\documentclass[reprint,aps,amsfonts,amsmath,amssymb,showpacs,showkeys,longbibliography,pra]{revtex4-1}
\usepackage{graphicx,array,natbib,subcaption,bm,dcolumn}
\usepackage{hyperref,amsbsy}
\usepackage{textgreek}
\usepackage[utf8]{inputenc}
\usepackage[T1]{fontenc}
\usepackage{mathptmx}
\DeclareMathOperator\erf{erf}

\begin{document}

\title{Numerical simulation of \textit{Ni}-like \textit{Xe} plasma dynamics and laser gain in a low inductivity capillary discharge}
\author{N.V. Kalinin}
\affiliation{Ioffe Institute, Russian Academy of Sciences, 26 Politekhnicheskaya street, Saint-Petersburg, 194021, Russia}
\author{R.M. Feshchenko},
\affiliation{P.N. Lebedev Physical Institute, Russian Academy of Sciences, 53 Leninski Prospekt, Moscow, 119991, Russia}
\email{rusl@sci.lebedev.ru}
\author{I.A. Artyukov},
\affiliation{P.N. Lebedev Physical Institute, Russian Academy of Sciences, 53 Leninski Prospekt, Moscow, 119991, Russia}
\author{V.A. Burtsev}
\affiliation{Ioffe Institute, Russian Academy of Sciences, 26 Politekhnicheskaya street, Saint-Petersburg, 194021, Russia}

\date{\today}

\begin{abstract}
X-ray lasers based on transitions in highly charged \textit{Ni}-like ions generating in the "water window" wavelength range can be pumped by compact laboratory discharge sources. This makes them promising candidates for use as compact coherent X-ray sources in laboratory applications including biological imaging and investigations of carbon containing materials. In this paper, the results of numerical simulations of the plasma dynamics and kinetics in an X-ray laser based on transitions in \textit{Ni}-like xenon ions are reported. The laser active medium is created by an extended low-inductive high current Z-discharge capable of producing two successive electrical pulses. The non-equilibrium multi-charged ion plasma dynamics is studied numerically using a non-stationary 1D two-temperature radiation-MHD model, which describes plasma hydrodynamics, non-stationary ionization, transfer of the continuum and line radiation as well as processes in the pumping electrical circuit. The ionic energy level populations are calculated in the quasi-stationary approximation. The simulation results allowed determination of the electrical and energy pumping parameters necessary to obtain a weak signal gain for the working transitions of the order of $g^+\sim1$ $\mbox{cm}^{-1}$. It was demonstrated that plasma with the electronic temperature of more than 400 eV and the density of more than $10^{19}$ $\mbox{cm}^{-3}$ can be created by a low inductive two step discharge with peak current exceeding 200 kA.
\end{abstract}


\maketitle

\section{Introduction} 
The tremendous progress in X-ray lasers over the last three decades has led to the demonstration of laser generation at numerous wavelengths ranging from 60.8 nm to 3.96 nm (in $Au^{51+}$ \textit{Ni}-like ions). The generation at such short wavelengths was achieved in hot laser-produced plasmas \cite{daido2002review} created at large-scale laser facilities, which were capable of providing optical pulse energies up to hundreds and thousands joules. In addition tabletop laser-pumped X-ray lasers generating short light pulses with the energy up to 100 {\textmu}J, wavelengths down to 10.9 nm and repetition rate up to 100 Hz have been demonstrated \cite{reagan2014high}. Moreover the laser gain was observed at even shorter wavelengths 6.85 nm and 5.85 nm \cite{wang2018compact}.

On the other hand, X-ray lasers pumped by the electrical discharge in capillaries have not advanced so much. In fact, the only operational device known is a capillary discharge X-ray laser (CDXL) working at the wavelength of 46.9 nm based on \textit{Ne}-like $Ar^{8+}$ ions \cite{vinogradov2003repetitively}. All attempts to create a shorter wavelength CDXL have not been successful so far \cite{kolacek2003principles,kolacek2008ways}. Despite the lack of progress the discharge pumped X-ray lasers (especially CDXLs) working at shorter wavelengths continue to attract considerable attention because they present a much more efficient way of converting the electrical energy into the short wavelength laser radiation as compared to optically pumped X -ray lasers, which should, finally, lead to generation of much higher average laser power and pulse energy. The discharge pumping efficiency is an important factor for possible future applications of compact X-ray lasers in such fields as X-ray microscopy in the "water window" (2.4--4.4 nm), interferometry of dense plasma, material sciences, nanotechnology, EUV lithography, etc \cite{Attwood2017Sakdinawat,kolacek2003principles,tan2007development}.

It should be mentioned that there are several ways to obtain a population inversion in hot plasmas \cite{elton2012x}. For instance, in the recombination pumped X-ray lasers the inversion can be obtained by recombination in a hydrodynamically cooled plasma with elevated population of highly excited hydrogen- or helium-like ions. On the other hand, the inversion in collisionally pumped X-ray lasers arises due to continues excitation of upper levels of a multi-charged ion by impacts of hot electrons, due to the cascading from higher-n states and by dielectronic recombination of ions with higher charges \cite{daido2002review, macgowan1990observation}. The lower level (the lower state) in the collisional scheme is quickly cleared by radiative decay to the ground state, the radiation from the upper level to the ground state being forbidden. The most favorable conditions for collisional excitation can be achieved in multi-charged ions with the electronic configurations similar to those of group VIII elements with closed electron shells, such as noble gases (e.g. \textit{Ne}-like ions) or transitional metals e.g. \textit{Ni}-like ions \cite{elton2012x}. The schemes based on \textit{Ni}-like ions usually provide lower gain but are more efficient in comparison with the schemes based on \textit{Ne}-like ions. The collisional pumping has so far been the only X-ray laser scheme used to achieve a significant laser generation \cite{daido2002review}. There are also schemes based on the optical pumping \cite{elton2012x}.

In principle, both recombination and collisional schemes can be realized in CDXLs although much attention has been historically focused at implementations of the recombination scheme in hydrogen-like nitrogen ions to achieve lasing at the technologically important wavelength of 13 nm \cite{kolacek2008ways}. Although as was already mentioned the only existing discharge pumped X-ray laser is based on the collisional scheme with \textit{Ne}-like $Ar^{8+}$ ions. The lack of progress in CDXLs can be explained by two factors. The first factor is the limited choice of active mediums because as a practical matter only gases work well as mediums in discharges, which limits the available options to noble gasses in \textit{Ne}-like and \textit{Ni}-like configurations. The second factor is the demanding requirements on the electrical pulse parameters especially the current rise rate, which needs to significantly exceed $10^{12}$ A/s in order to achieve a rapid detachment of the plasma from the capillary walls and to heat it to a high electron temperature necessary to produce and excite highly charged \textit{Ne}-like and \textit{Ni}-like ions required for the short wavelength lasing \cite{kolacek2008ways}. It is also desirable to limit the required peak current and voltage to prolong the life of the capillary \cite{tan2007development}. Realizing such fast discharges necessitates designing novel low-inductance voltage generators and electrical current transmission lines.

In this work, we report the results of numerical simulations of a high current capillary discharge plasma of xenon, which was used before in laser driven pumping schemes to produced plasmas with the gain coefficient reaching as high as 17.4 $\mbox{cm}^{-1}$ \cite{lu2002demonstration}. The xenon plasmas are produced by fast electrical pulses of low-inductance voltage generator with the forming lines described earlier \cite{burtsev2013matching} and contained \textit{Ni}-like ions. A two stage pumping scheme is considered where the plasma is created by a pre-pulse formed by the same generator. We demonstrate that a gain exceeding $g^+\sim1$ $\mbox{cm}^{-1}$ can be obtained by a suitable choice of gas pressure, electrical and capillary parameters. Such a value of the gain coefficient in broadly inline with the measured gains in discharge plasmas as reported before \cite{kolacek2003principles}. 

\section{Model of capillary discharge plasma}
The numerical simulation of the active medium of a short wavelength laser based on plasma with multi-charged \textit{Ni}-like \textit{Xe} ions was carried out in several interconnected stages following the same principles as described earlier for the nitrogen and \textit{Ne}-like argon discharge plasmas \cite{burtsev2014heating,burtsev2017numerical}. 

At the first stage, the ionization, heating and dynamics of the plasma were modeled. An 1D axisymmetric two-temperature (2T)  radiation magnetohydrodynamics (MRHD) model was used. It allowed us to calculate the spatio-temporal characteristics of the multi-charged \textit{Xe} ion plasma e.g. the temperature and concentration of ions and electrons, the ionic composition as well as the spatial distribution and temporal evolution of electrical and magnetic fields and temporal evolution of the electrical pulse. The model also accounted for the line and continuum radiation transfer.

In the second stage the atomic kinetic modeling of working level population densities of ions, line intensities and the calculation of gain coefficient was conducted in the quasi-stationary approximation. The MRHD simulations of the plasma heating and dynamics (the first stage) and calculations of the atomic level population kinetics (the second stage) were performed independently because the electronic excitation and decay of ionic levels do not effect the plasma dynamics significantly due to the vastly different time scales. 

The magnetohydrodynamics of plasma was modelled using the following system of 1D axisymmetric equations written through the Lagrangian variables in the SI system of units
\begin{align}
\frac{du}{dt}=&-r\frac{\partial P}{\partial m}+F,\label{1a1}\\
P_\omega=&-\rho(\nu_i+\nu_e)\frac{\partial u}{\partial m}+\mu_a\rho\left(\frac{\partial u}{\partial m}\right)^2,\label{1a2}\\
F=&-\frac{\rho}{\mu_0r}\frac{\partial}{\partial m}\left(rB_\varphi\right)^2,\label{1a3}\\
\frac{d\varepsilon_e}{dt}=&P_e\frac{d}{dt}\left(\frac{1}{\rho}\right)-\frac{\partial W_e}{\partial m}+\nonumber\\
&Q_J+Q_{ei}-Q_{rad}-\frac{d\varepsilon_{ion}}{dt},\label{1a4}\\
\frac{d\varepsilon_i}{dt}=&P_i\frac{d}{dt}\left(\frac{1}{\rho}\right)-Q_{ei}-P_\omega\frac{d}{dt}\left(\frac{1}{\rho}\right),\label{1a5}\\
W_e=&-\chi_e\rho r^2\frac{\partial T_e}{\partial m},\label{1a6}\\
Q_J=&\frac{1}{\mu_0\sigma}E_z\frac{\partial}{\partial m}\left(rB_\varphi\right),\label{1a7}\\
\frac{d}{dr}\left(\frac{B_\varphi}{r\rho}\right)=&-\frac{\partial E_z}{\partial m},\label{1a8}\\
E_z=&\frac{\rho}{\mu_0\sigma}{\partial m}\left(rB_\varphi\right),\label{1a9}\\
\frac{\partial}{\partial t}\left(\frac{1}{\rho}\right)=&\frac{\partial u}{\partial m},\label{1a10}
\end{align}
where $t$ is the time, $r$ is the radius, $z$ is the coordinate along the plasma column, $\varphi$ is the azimuthal angle, $u=dr/dt$ is the radial velocity, $\rho=(dm/dr)/r$ is the density of plasma, $m$ is the plasma mass within radius $r$ per a unit length in $z$ direction and per a unit of azimuthal angle (the Lagrangian variable here), $P_i$, $P_e$, $P_\omega$ and $P=P_i+P_e+P_\omega$ are the ionic, electron, viscose and total pressure, respectively. The viscous pressure defined by \eqref{1a2} includes two terms: the real viscosity with coefficients $\nu_i$, $\nu_e$ and the effective viscosity with coefficient $\mu_a$. Parameters $\varepsilon_i$, $\varepsilon_e$ and $\varepsilon_{ion}$ are the ionic, electron and ionization energy densities, respectively. Also $T_e$ is the electronic temperature, parameter $Q_{ei}$ is the specific energy flux from ions to electrons due to elastic collisions, $Q_J$ is the specific Joule heating flux, $\chi_e$ is the electronic thermal conductivity coefficient, $\sigma$ is the electrical conductivity of plasma, $E_z$ is the longitudinal component of electrical field, $Q_{rad}$ is the specific radiation flux from plasma and $B_\varphi$ is the azimuthal component of magnetic field. The ionic energy levels and other information were taken from the NIST and other databases \cite{ivanova2015wavelengths,ivanova2018features,kelleher1999new}.

\begin{figure}
\includegraphics[scale=0.45]{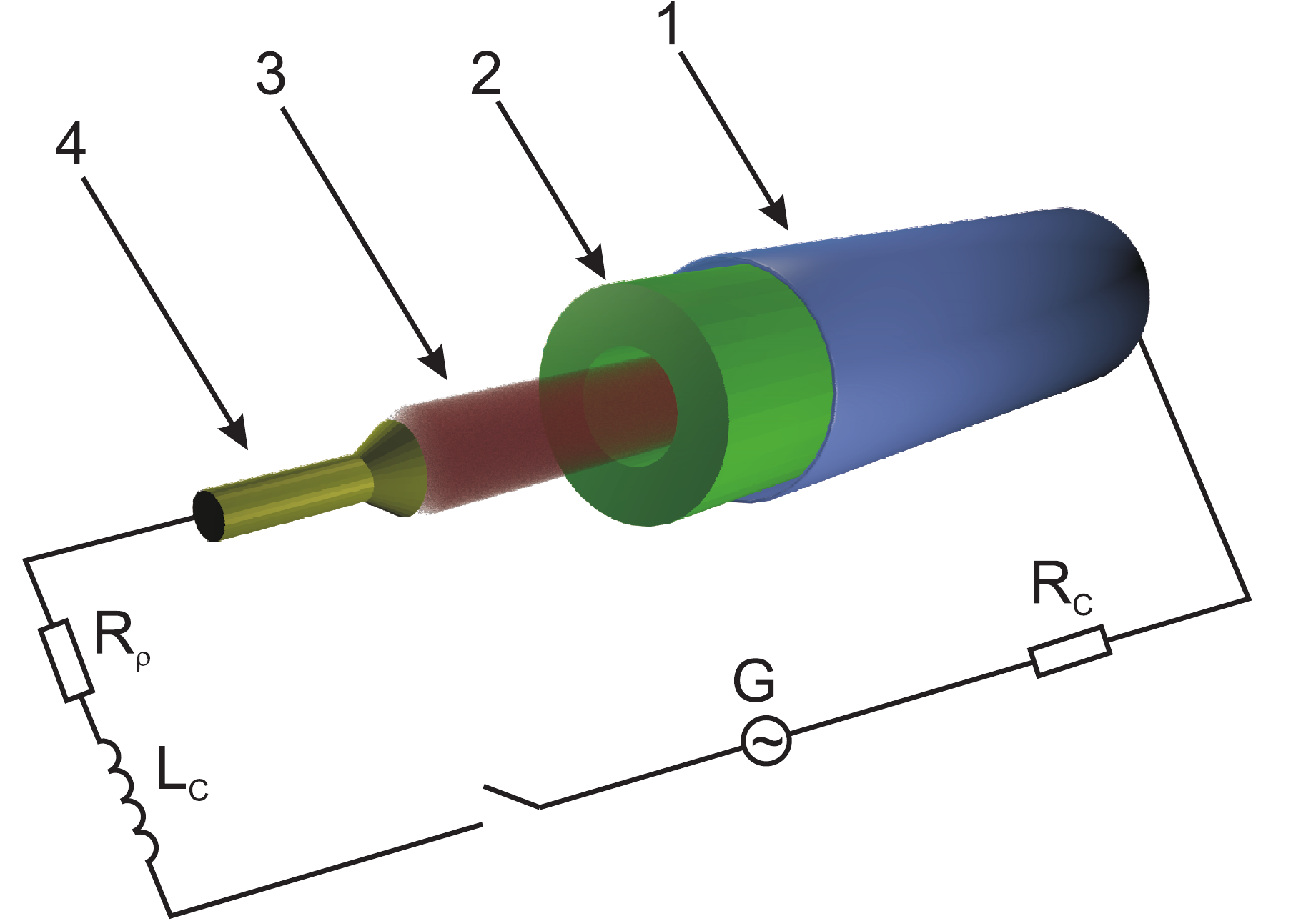}
\caption{Electrical schematic diagram of the discharge circuit: G --  generator of high voltage pulses, 1 --  reverse current line, 2 --  capillary, 3 -- plasma and 4 --  high-voltage electrode. (Color online) }
\label{f1}
\end{figure}

The magnetic field at the boundary of plasma column depends on flowing current $I$ and voltage $U$, which are determined from the electric circuit equation 
\begin{equation}
U_G(t)-(R_\rho+R_C)I-L_C\frac{dI}{dt}-\frac{d}{dt}L_{\Delta}I=0,
\label{1b}
\end{equation}
where $U_G(t)$ is the voltage supplied by generator G, $R_\rho$ -- the wave impedance of transmission line, $L_C$ and $R_C$ are the inductance and impedance (variable), respectively, of the discharge electrical circuit and $L_\Delta$ is the inductance of the gap between the external surface of plasma column and the reverse current line. The schematic electrical diagram of the discharge circuit is shown in FIG.~\ref{f1}.

The equations of state for shell and column plasmas are based on the average ion approximation. To calculate the transfer coefficients we used an empirical model valid for a broad range of parameters \cite{shlyaptsev1994modeling,shlyaptsev2003numerical}.

The ionic composition of plasma is determined by the following equation \cite{vinogradov1987characteristics}
\begin{equation}
\frac{1}{N_e}\frac{d\alpha_Z}{dt}=\alpha_{Z}I_{Z-1}-\alpha_Z(I_{Z}+R_Z)+\alpha_{Z+1}R_{Z+1},
\label{1c}
\end{equation}
where $N_e$ is the concentration of  electrons, $I_Z$ and $R_Z$ are the ionization and recombination rates for ions with charge $z=1,2,...,Z_n$ and $Z_n$ is the nucleus charge. The relative concentration of ions with charge $Z$ is defined as
\begin{equation}
\alpha_Z=\frac{\sum\limits_{\kappa}N_{\kappa Z}}{\sum\limits_{Z}\sum\limits_{\kappa} N_{\kappa Z}}=\frac{\sum\limits_{\kappa}N_{\kappa Z}}{N},
\label{1d}
\end{equation}
where $N$ is the total concentration of ions, $N_{\kappa Z}$ is the concentration of ions with charge $Z$ in excited state $\kappa$. Equation \eqref{1c} takes into account the ionization by electron impacts, photo- and three-body recombination to the excited and ground states as well as the dielectronic recombination.

The model also accounts for the plasma cooling by Bremsstrahlung, recombination, dielectronic and line radiation \cite{vinogradov1983amplification, vinogradov1987characteristics,derzhiev1986radiation}
\begin{equation}
Q_{rad}=Q_B+Q_R+Q_D+Q_L,
\label{1d1}
\end{equation} 
where the power of Bremsstrahlung radiation $Q_B$, recombination radiation $Q_R$ and dielectronic recombination radiation $Q_D$ were calculated using the following empirical formulas
\begin{align}
Q_B=&1.6\times10^{-32}\sqrt{T_e}\sum\limits_{Z}\alpha_{Z}Z_{eff}^2,\label{1e1}\\
Q_R=&1.6\times10^{-19}\sum\limits_{Z}\alpha_{Z}R_Z\left(I_{Z-1}+\frac{3}{2}T_e\right),\label{1e2}\\
Q_D=&1.6\times10^{-19}\times\nonumber\\
&\sum\limits_{Z}\alpha_Z\sum\limits_{\kappa} R_{\kappa Z}(\Delta E+I_Z-E_n),\label{1e3}
\end{align}
where $Z_{eff}$ is the effective ion charge, $\Delta E$ is the transition energy and the $E_n$ is the energy of the n-th electronic excited state.

The line radiation was calculated taking into account only transitions to the ground state with $\Delta n=0,1$ through the following empirical formula
\begin{equation}
Q_L=1.6\times10^{-19}\sum\limits_{Z}\alpha_Z\sum\limits_{n=0,1}\frac{I_{0n}\Delta E_{0n}}{1+N_e I_{n0}/A_{n}},
\label{1f}
\end{equation}
where $A_n$ is the spontaneous emission rate from state $n$ to the ground state $n=0$, $\Delta E_n=E_n-E_0$ is the transition energy and $I_{0n}$ is the collisional transition rate from the n-th state to the ground state.

\begin{figure}
\begin{subfigure}{0.51\textwidth}
 \includegraphics[width=1\linewidth]{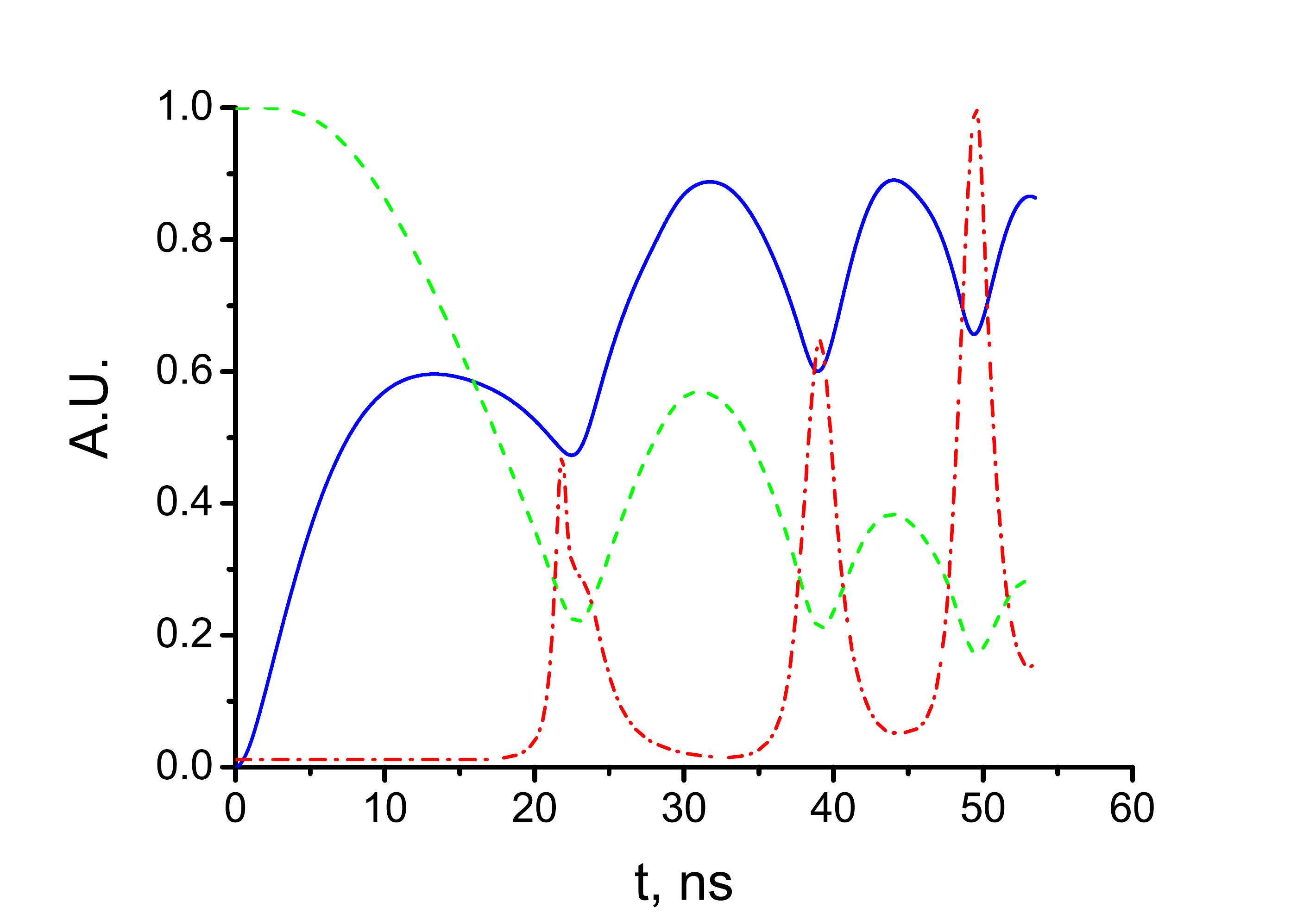}\caption{}\label{f2a}
\end{subfigure}
\begin{subfigure}{0.51\textwidth}
 \includegraphics[width=1\linewidth]{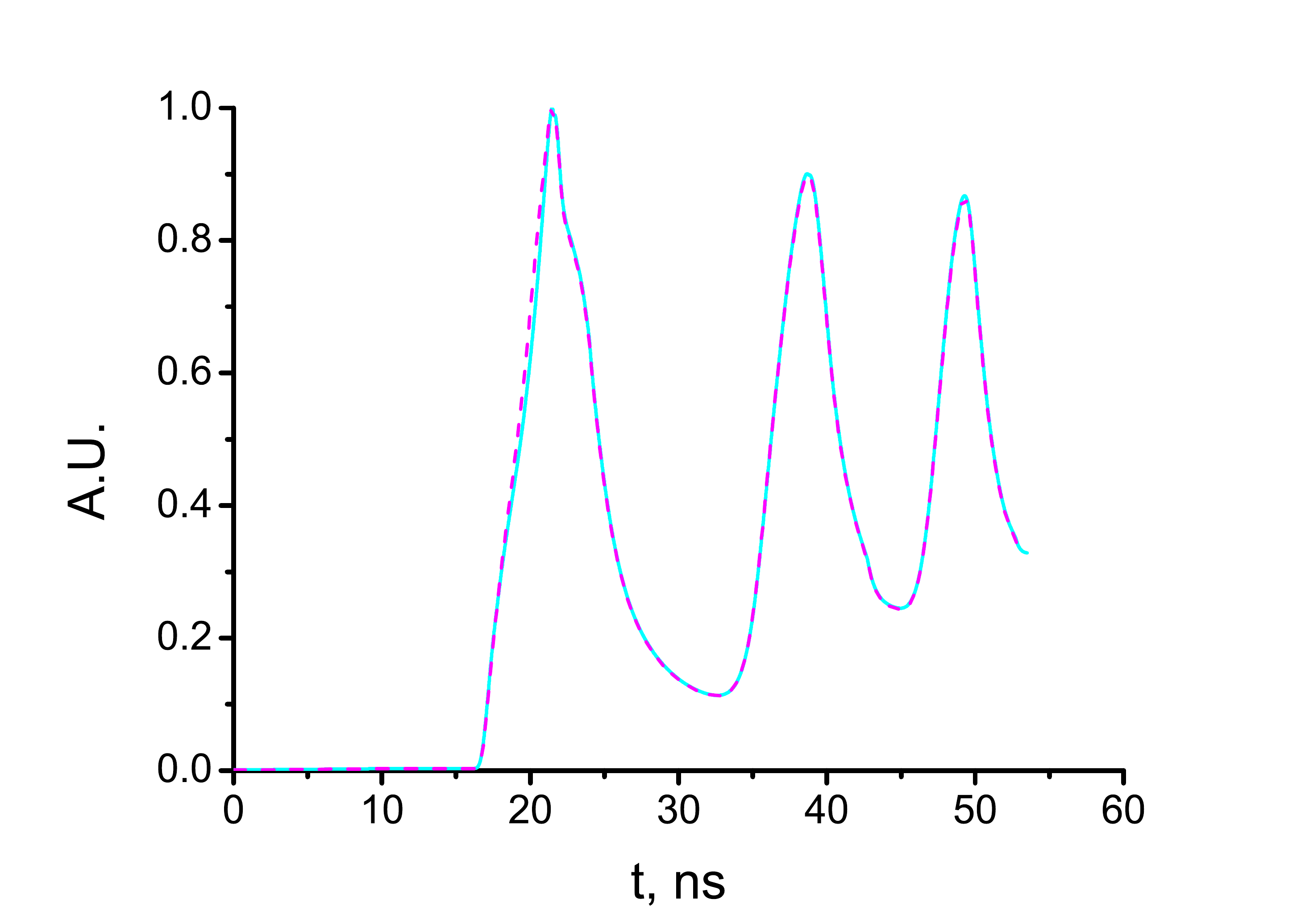}\caption{}\label{f2b}
\end{subfigure}
\caption{Temporal dependencies of (a): the discharge current $I$ -- blue solid line, external plasma column radius $r_{ext}$ -- green dash line and plasma density at the discharge axis $\rho_c$ -- red dash-dot line; (b): electronic temperature $T_e$ -- cyan solid line and ion temperature $T_i$ -- magenta dash line. All parameters were normalized to their maximum values.(Color online)}
\label{f2}
\end{figure}

\begin{figure}
\includegraphics[width=1\linewidth]{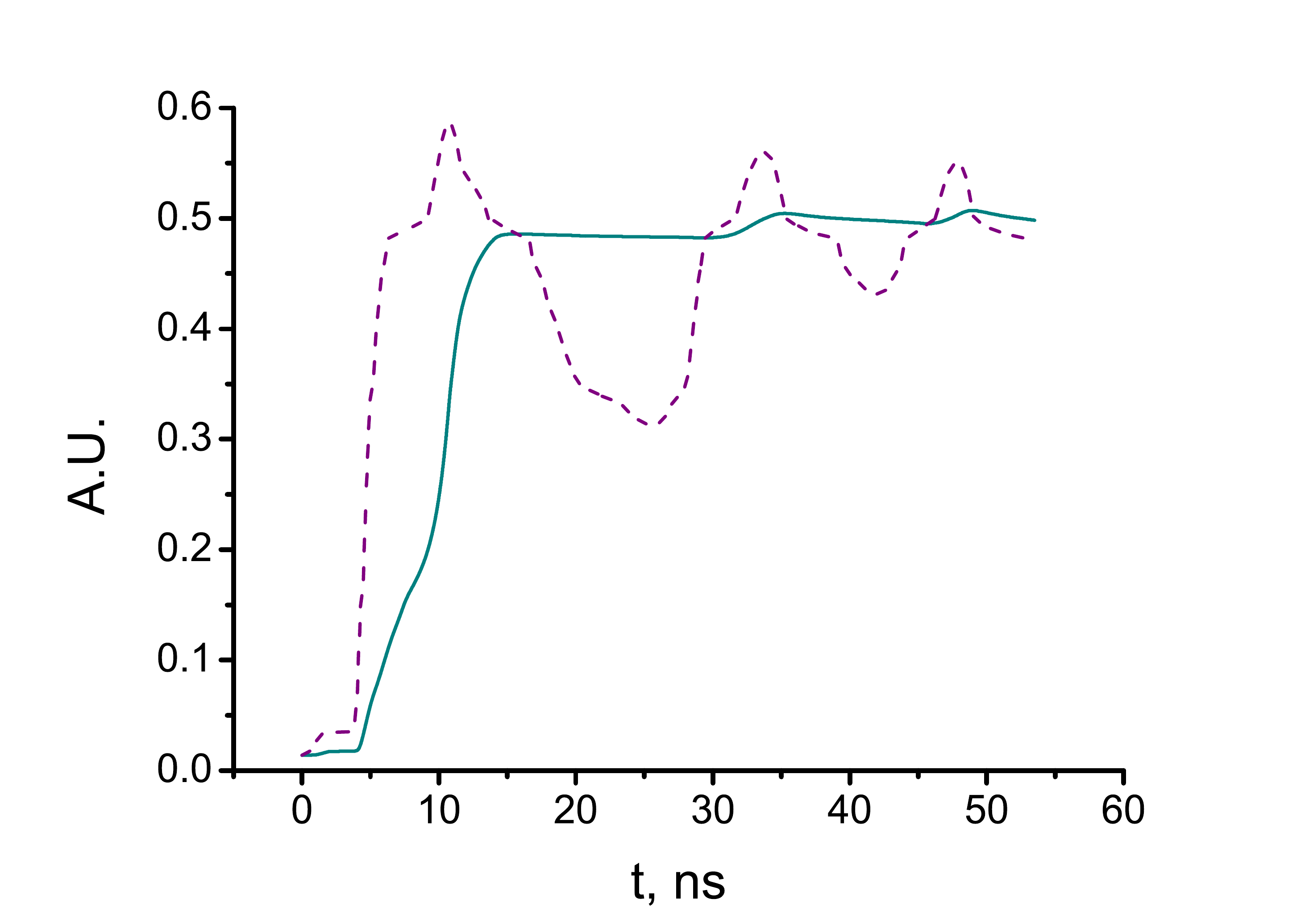}
\caption{Temporal dependencies of  equilibrium average ion charge $Z_a$ at the discharge's axis -- dark cyan solid line and non-equilibrium average ion charge $Z_a$ at the discharge's axis -- purple dash line. The ion charges were normalized to nucleus charge $Z_n = 54$.(Color online)}
\label{f2c}
\end{figure}

The radiative-collisional model of plasma of \textit{Ni}-like ions is similar to that of \textit{Ne}-like ions. The upper working energy level 3d4d[J=0] is populated by collisions with hot electrons and decays radiatively only to the lower working level 3d4p[J=1], the radiative decay to the ground state being forbidden. The lower working level is cleared very fast by radiative decays to the ground state. The cascading from higher-n states as well as dielectronic recombination of Co-like ions \cite{macgowan1990observation} were both neglected at this stage, as our estimate showed that it had a negligible effect on the gain coefficient.

As was mentioned above, the second stage of simulations included calculations of population densities $N_{\kappa Z}$ of ions in excited state $\kappa$ with charge $Z$ in the quasi-stationary approximation. All population densities were considered to be dependent on the local electron temperature, plasma density and ionic composition \cite{vinogradov1983amplification}
\begin{align}
\frac{dN_{\kappa Z}}{dt}=&0,\label{1g1}\\
\sum\limits_{i}N_{iZ}S_{\kappa Z}=&N_{\kappa Z}\sum\limits_{i}S_{\kappa iZ},\label{1g2}
\end{align}
where $S_{\kappa iZ}$ is the total (radiation and impacts) probability of transition from  level $i$ to level $\kappa$ of ions with charge $Z$.  As was noted above this can be justified by vastly different time scales of ionization and population level dynamics: the former is much slower than the latter, which timescales decrease with the ion charge as $Z^{-4}$ or $Z^{-5}$.

\begin{figure}
\includegraphics[scale=0.33]{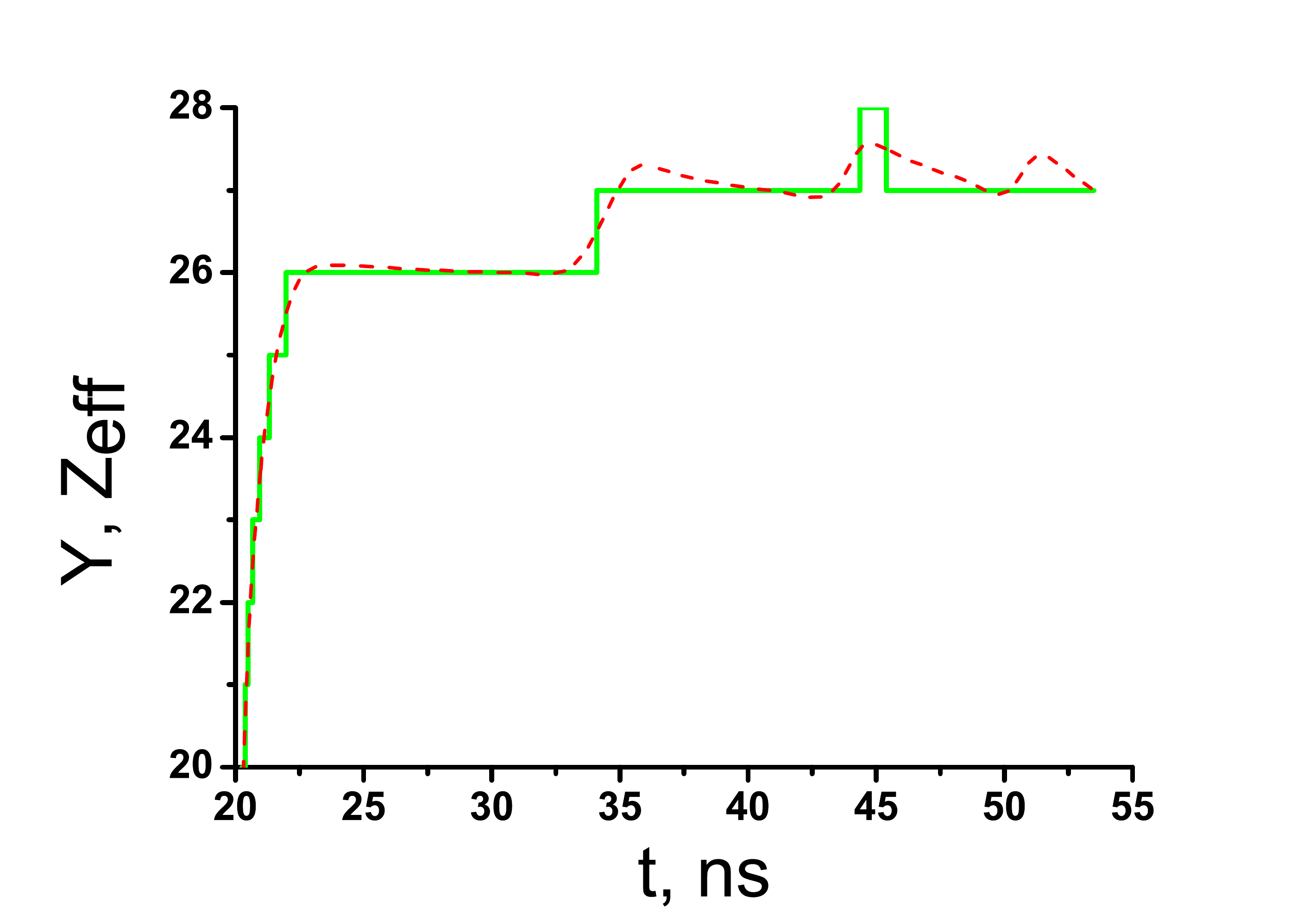}
\caption{Temporal dependencies of the average ion charge ($Z_a$, red dash line) and median ion charge ($Y$, green solid line) on the axis of the capillary discharge as function of time.(Color online)}
\label{f3}
\end{figure}

Finally, population inversion $\Delta N$ and the gain in the emission line center were calculated using the following formula 
\begin{equation}
g^+=\frac{\lambda^2A\Delta N}{4\sqrt{\pi}\Delta\omega},
\label{1h}
\end{equation} 
where $A$ is the spontaneous emission rate from upper working state to lower working state, $\lambda$ is the central wavelength and $\Delta\omega$ is the spectral width of working radiative transition. 

The working level population inversion and gain depend on the resonant radiation trapping, which is determined by the optical thickness of plasma in the transverse direction. This effect is usually accounted for with the Biberman--Holstein method where resonant transition probabilities $A$ are replaced with effective probabilities $A_s$ \cite{golubovskii2017radial}, which depend on the radiative volume escape factor (See Appendix \ref{apA} for details). This factor in turn depends on the geometry of the plasma volume, spectral function shape, lower level population and on the optical thickness of the plasma. The approximate escape factors used here were taken from \cite{derzhiev1986radiation}.
	
To solve numerically the system of the plasma dynamics equations \eqref{1a1}--\eqref{1a10},  electrical circuit equation \eqref{1b}, radiation transfer equations \eqref{1d1}--\eqref{1f}, ionic population equations \eqref{1c} and level kinetic equations \eqref{1g1}--\eqref{1g2} we used a conservative implicit finite-difference scheme. The computations were conducted step-by-step by splitting them into different physical processes with different timescales as well as by combination marching \cite{samarsky2004popov}.

\section{Simulation results}
In papers \cite{shlyaptsev1994modeling,shlyaptsev2003numerical} the parameters of \textit{Ni}-like \textit{Xe} plasma needed for the amplification of spontaneous X-ray radiation were found. For transitions 4f--4d (2--1), 4d--4p (3--2), 4d--4p (0--1) they are  $T_e>400$ eV and $N_e\approx5\times10^{20}$ $\mbox{cm}^{-3}$. In the present calculations we followed these works but varied the gas pressure, diameter of the capillary, rise time of the electrical pulse and voltage amplitude. 

\begin{figure}
\begin{subfigure}{0.51\textwidth}
 \includegraphics[width=1\linewidth]{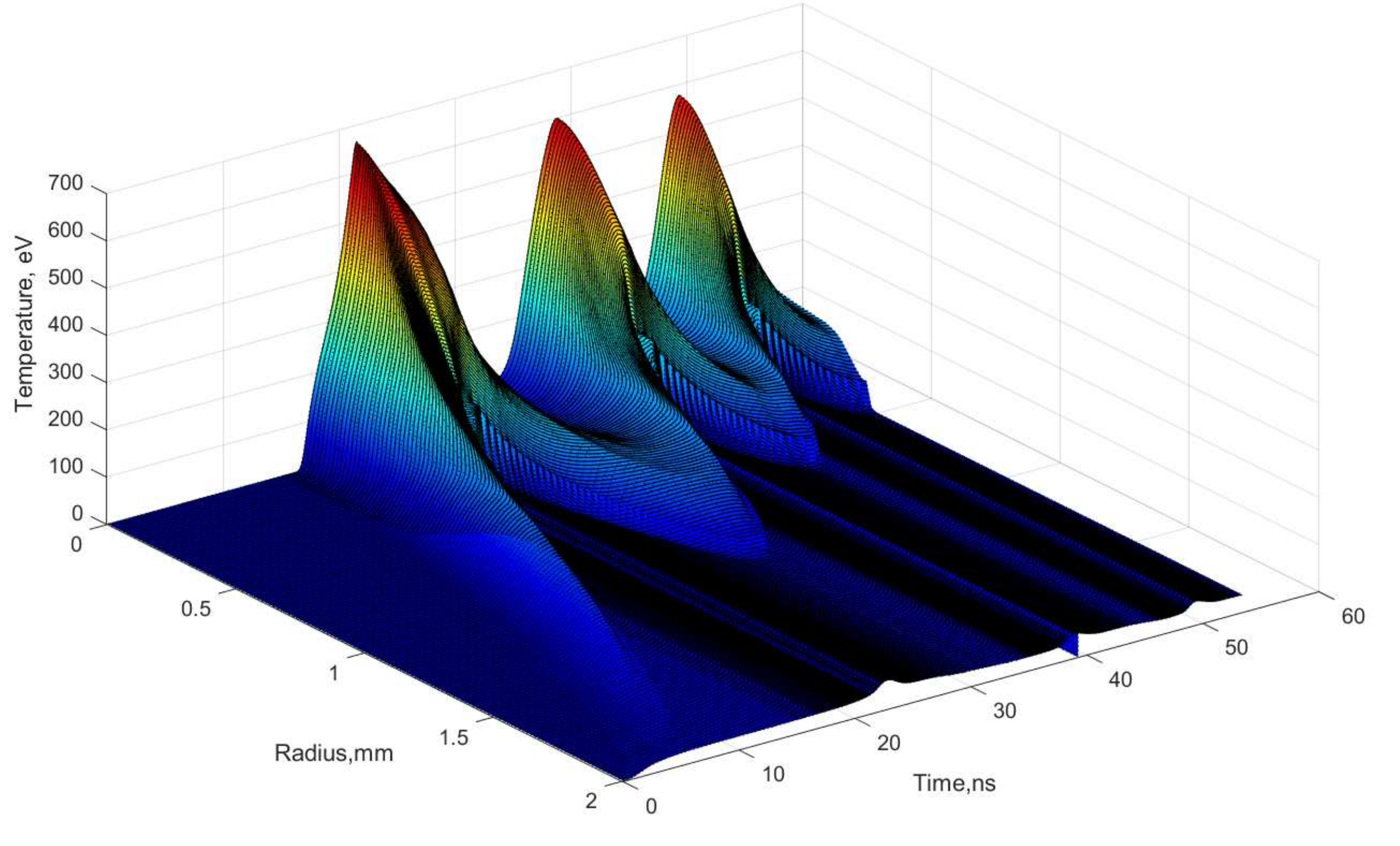}\caption{}\label{f4a}
\end{subfigure}
\begin{subfigure}{0.51\textwidth}
 \includegraphics[width=1\linewidth]{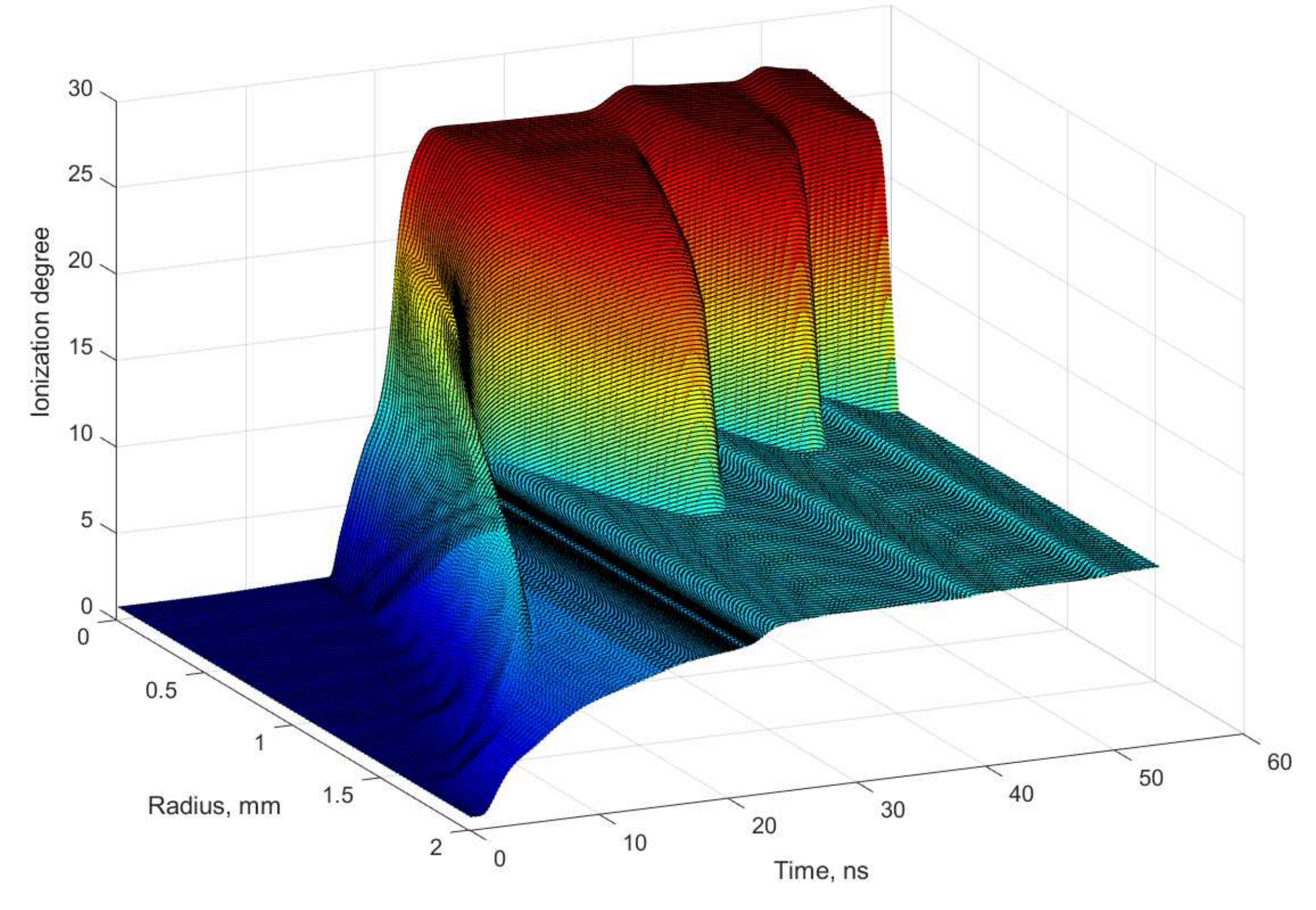}\caption{}\label{f4b}
\end{subfigure}
\caption{Dependencies of the normalized electronic temperature $T_e$ (a) and the average ion charge $Z_a$ (b) on the time and on the plasma column radius.(Color online)}
\label{f4}
\end{figure}

The simulations were done taking into account the characteristics of the Extreme-M experimental setup \cite{burtsev2013matching} developed in Ioffe Physical-Technical Institute (See photo in Supplemental Material at [URL will be inserted by publisher]). The pulse voltage rise time was 1--2 ns, the electrical pulse duration and amplitude were set at 100 ns and 200 kV, respectively, and wave resistance $R_\rho$ was assumed to be 1 \textOmega. 

In FIG.~\ref{f2a} the temporal dependencies of the discharge current (blue solid line) and external diameter of \textit{Xe} plasma column (green dash line) are depicted. The initial gas pressure was 4 Torr and the length of the discharge chamber was assumed to be 10 cm. The power supply and pulse plasma load were matched. The first current rise time (about 12 ns) was shorter than the first plasma compression time (about 22 ns). It resulted in a highly efficient conversion of the magnetic energy into the energy of plasma, that is relatively common for high current discharges. 

After reaching its maximum of about 150 kA the current (see FIG.~\ref{f2a}) decreased to about 120 kA in 10 ns and then increased again to about 220 kA and after that remained relatively stable despite the significant reduction of the external plasma column radius and the later increase of the inductance between the plasma column and reverse line. This behavior of the current differed significantly from what would occur in the case of a discharge characterized by a sinusoidal temporal dependence of the current.

In FIG.~\ref{f2a} one can also see that the external plasma radius (see green dash line) decreased to about 1/5-th of its initial value at the end of the first compression stage at about 22 ns. Then the plasma expanded for about 8 ns towards about 3/5-th of the initial radius and contracted again down to 1/5-th of the initial radius at the secondary compression stage. Finally, the plasma radius did not change considerably for a period of about 15 ns and stayed within the 1/5--1/3-th of the initial radius. 

The part of plasma where the radiation amplification takes place is concentrated near the axis of the discharge and has a typical diameter of several tens of nanometers with sufficient homogeneity in the radial direction. The ion charge distribution in the plasma is known to be rather steep, with the most ions being either median one with charge $Y$ or two its closest neighbors with charges $Y-1$ and $Y+1$. Therefore, it is quite reasonable to consider the temporal evolution of this population of median ions with charge $Y\approx Z_a$. As one can observe in FIG.~\ref{f2c}, in the equilibrium plasma model (dark cyan solid line) the average ion charge $Z_a$ followed all the changes of the plasma temperature and density but in the case of the non-equilibrium model (purple dash line) it lagged behind the plasma temperature and density. 

On the other hand, at the end of the first compression stage at 22 ns the average ion charge $Z_a$ reached 26 (see FIG.~\ref{f3}), while the electronic temperature $T_e$ went to about 690 eV (see FIG.~\ref{f2b}, cyan solid line) and the plasma compression to about 50. On the other hand the electron density reached about $6.6\times10^{18}$ $\mbox{cm}^{-3}$ while the median ion charge $Y$ became 26 and remained constant until about 34 ns (see FIG.~\ref{f3}). After that the plasma cooled down for about 12 ns to $T_e\approx 60$ eV due to its expansion and the strong line radiation. During this cooling the equilibrium median ion charge $Y$ remained stable at 26. During the second compression stage after about 34 ns $Y$ slightly increased to 27 for about 3 ns with the electronic temperature going up to about 600 eV at 39 ns. The maximum electronic temperature, plasma density and compression in the secondary compression stage reached about 590 eV, $10^{19}$ $\mbox{cm}^{-3}$ (see FIG.~\ref{f2a}, red dash-dot line and FIG.~\ref{f3}) and 100, respectively, but the median charge $Y$ still remained at 27.

The radial and temporal distributions of the electronic temperature and average ion charge are shown in FIG.~\ref{f4}. These graphics demonstrate the dynamics of plasma as well as possibility of obtaining plasma parameters to get a gain in excess of 1 $\mbox{cm}^{-1}$. 

\begin{figure}
\includegraphics[scale=0.33]{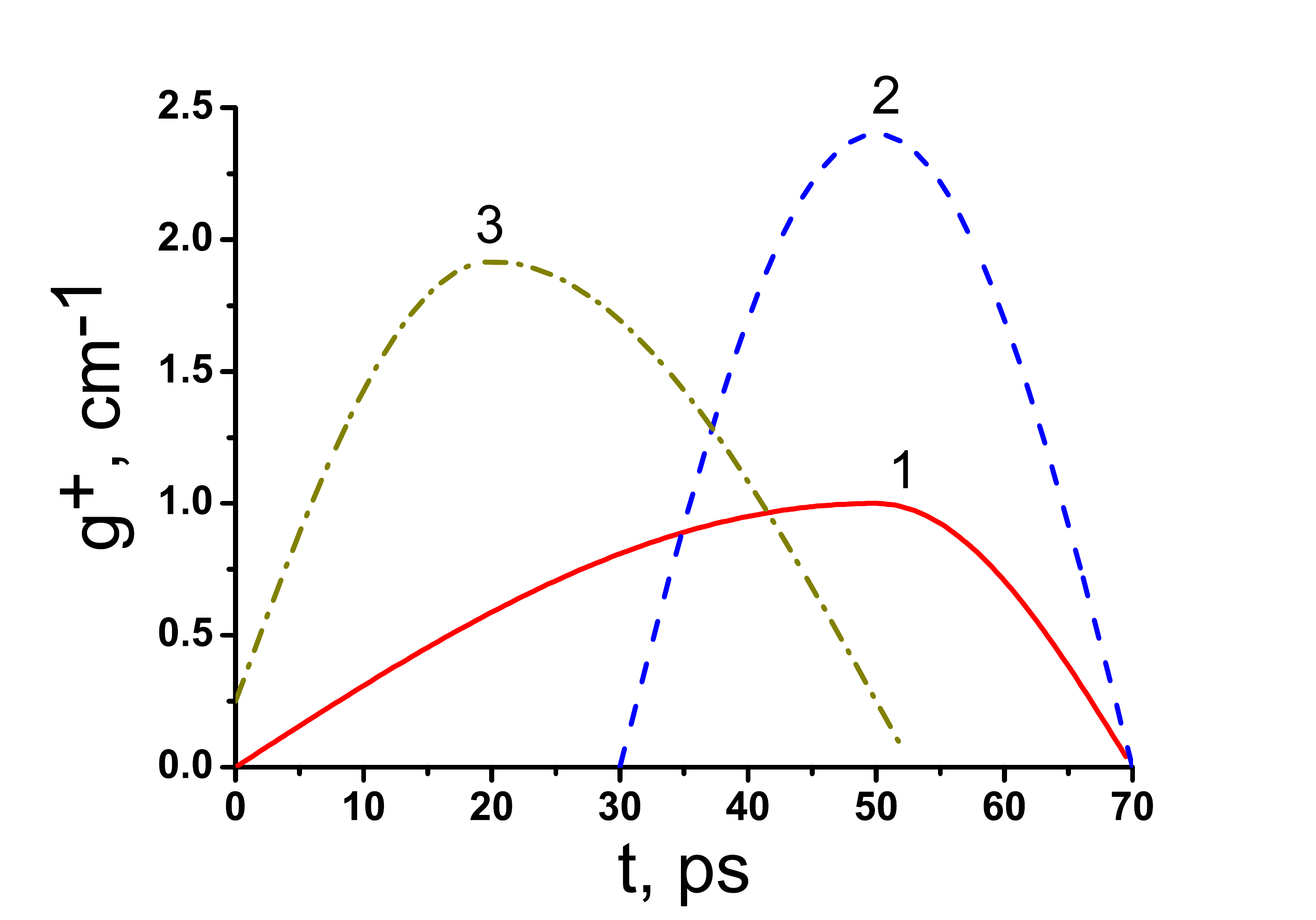}
\caption{Temporal dependencies of the gain coefficient $g^+$ in the center of capillary discharge for the internal capillary diameter of 4 mm and the following transitions in \textit{Ni}-like \textit{Xe} ions: 1 -- 4f--4p ($\lambda=13.2$ nm),  2 -- 4d--4p ($\lambda=9.9$ nm), 3 -- 4p--4s ($\lambda=17.3$ nm). The time is measured from the moment when the median ion charge reaches the target value of $Y=27$.(Color online)}
\label{f5}
\end{figure}

The calculated gains $g^+$ for three different transitions are depicted in FIG.~\ref{f5}. One can see a good qualitative agreement of the presented calculation results with the values obtained for the laser pumped X-ray lasers in work \cite{ivanova2001theoretical}. However, the maximum gain coefficients turned out to be much lower than those reported in \cite{lu2002demonstration,ivanova2001theoretical}, which can be explained by the much lower density of the discharge plasma as compared to the laser plasma. FIG.~\ref{f4} and \ref{f5} show that the maximal gain coefficient rises with increase of the electronic temperature. The gain coefficients exceeded 1 $\mbox{cm}^{-1}$ for a few picoseconds. It should be of interest to calculate the gain for different capillary and plasma column diameters. For example, in the case of a high electron density the calculations have demonstrated a decline of the gain as the plasma column diameter increases. On the other hand, if the electron density is low, the gain only weakly depends on the plasma column diameter.
    
\section{Conclusions}

The reported numerical simulations of a \textit{Ni}-like \textit{Xe} plasma created by a high current low-inductive discharge point to a feasibility of an X-ray laser source working at a wavelength near 10 nm with the gain exceeding 1 $\mbox{cm}^{-1}$. Two-stage (pre-pulse and main pulse) active medium formation was demonstrated using the power supply based on electrical forming lines with distributed parameters. This pumping scheme is an analog of the well-known two-pulse X-ray laser pumping scheme. The simulation results indicated that a plasma with the electronic temperature of more than 400 eV and the density of more than $10^{19}$ $\mbox{cm}^{-3}$ can be created by the low-inductive two-stage discharge with the peak current exceeding 200 kA.

The radiation magnetohydrodynamic (MRHD) model used in the simulations is based on the standard two-temperature 1D approximation, which takes into account the plasma hydrodynamics, magnetic field evolution, radiation transfer by both continuum (Bremsstrahlung, recombination and di-electronic) radiation and line radiation. On the other hand, the ionic level populations were estimated in the quasi-stationary approximation assuming that the corresponding processes are much faster than the MRHD timescales. The gain coefficient was calculated using the population densities of the ions in working states with a simple formula. The resonant radiation trapping was also accounted for using the Biberman--Holstein method. The simulations demonstrated that the gain exceeding 1 $\mbox{cm}^{-1}$ can exist for a few picoseconds for two transitions of \textit{Ni}-like \textit{Xe} ions: 4d--4p at 9.9 nm and 4p--4s at 17.3 nm when the gas pressure is about 4 Torr and the capillary diameter is about 4 mm. This value is typical for discharge pumped X-ray plasmas, where the measured weak signal gain coefficient ranges from 0.6 to about 3 $\mbox{cm}^{-1}$ \cite{kolacek2003principles}. The lower gain in discharge plasmas is compensated by a large length of the discharge column, which can exceed 10 cm.

The future improved models of the discharge plasmas will utilize a more detailed simulation of the kinetics of ionic levels and of the weak signal gain using more realistic spectral, collisional, energetic and optical characteristics of ions. This will allow us to better account for the re-absorption of the radiation, excitation of the working levels e.g. by taking into account the cascading from higher-n states and by the dielectronic recombination of Co-like ions, which may either increase or diminish the gain coefficient. We are also planning to study hydrodynamic instabilities, which inevitably arise during the multistage plasma compression and expansion. We strongly believe that an optimized electrical discharge with sufficiently short current rise time can provide for an efficient excitation of the upper working levels of Ni-like ions especially after all possible excitation channels are properly accounted for in those future improved plasma models. 

\section*{Supplementary Material}
See Supplementary Material for a photo the Extreme-M experimental setup.

\begin{acknowledgments}
The authors would like to thank A.V. Vinogradov for the fruitful discussions. The work was supported by the Research Programme of the Presidium of the Russian Academy of Sciences \textit{Actual problems of photonics, probing inhomogeneous mediums and materials} PP RAS No 7 as well as by the basic funding within the framework No 0023-0002-2018 as well as under RFBR grant No 10-08-01066.
\end{acknowledgments}

\appendix
\section{Accounting for radiation trapping by the Biberman--Holstein method}
\label{apA}
To take into account induced transitions the Biberman--Holstein approximation is usually used, which is based on the assumption that plasma characteristics are only weakly inhomogeneous. In this approximation a linear escape factor $\Theta$ is introduced, which relates the spontaneous radiation transition rates to their effective values $A'_{mm'}=\Theta_{mm'}A_{mm'}$. The escape factor depends on geometry of the plasma, its optical thickness, the shape of spectral function and on the population of the lower level.  

In the present case the plasma has a cylindrical shape and is characterized by sharp inhomogeneities in the radial direction. In such a situation accounting for the re-absorption in this cylindrical plasma column is generally not possible by introducing only a local escape factor with the spectral function independent of the radius. On the other hand, one of the notable features of capillary plasma is formation of a dense kernel at the longitudinal discharge axis, where the plasma density is sufficiently homogeneous, which justifies the use of the Biberman--Holstein approximation in the vicinity of plasma axis. 

The fast macroscopic movements of the discharge plasma make the dynamical Doppler effect important. In the approximate automodelic description of plasma motions with boundary velocity $u_0$ for the Doppler line shape $\Theta_D$ the following formula has been obtained
\begin{multline}
\Theta_D\approx\frac{2\delta}{\sqrt{\pi\eta}}\times\\
\left[1-\left(\sqrt{\pi}\left(1-\erf(\delta+\sqrt{\ln\eta})\right)-\frac{\sqrt{\pi}\eta}{2\delta}\erf(\delta)\right)\right],
\label{ap1}
\end{multline}
where $\eta=k_0r_0$ is the optical thickness, $k_0$ is the absorption coefficient at the center of the line, $r_0$ is the kernel radius, $\delta=\gamma_0/\gamma_D$, $gamma_D$ is a characteristic Doppler width and $\gamma_0=\omega_0 u_0/c$ is characteristic length due to dynamical Doppler effect. In the limiting case $\eta\gg1$ and $\delta\to0$ (\ref{ap1}) can be re-written as 
\begin{equation}
\Theta_D(\eta)\approx\frac{\pi}{4\eta}\sqrt{\pi\ln\eta}.
\label{ap2}
\end{equation}
Finally, due to weak dependence of of the escape factor on the plasma geometry we have for arbitrary points within radius $r$ the following equation for $\Theta$
\begin{equation}
\Theta_D\approx\frac{1}{2}\left(\Theta(r-r_0)+\Theta(r+r_0)\right).
\label{ap3}
\end{equation}

\bibliography{XeKrcapillaryLaserPhysRevA_arxiv}

\end{document}